\documentclass[aps,twocolumn,nofootinbib,superscriptaddress]{revtex4-2}
\pdfoutput=1
\usepackage{graphicx}  
\usepackage{dcolumn}   
\usepackage{bm}        
\usepackage{amsmath,amssymb,amsxtra,bm,epsfig}   
\usepackage{float}
\usepackage[dvipsnames]{xcolor}
\usepackage{xspace}
\usepackage{hyperref}
\usepackage{orcidlink} 
\usepackage{cancel,soul,ulem}
\hypersetup{
	colorlinks=true,
	linkcolor=red,
	filecolor=magenta,
	urlcolor=ForestGreen,
	citecolor=RoyalBlue}

\usepackage[]{caption}
\newcommand{\bmt}{\begin{pmatrix}}
	\newcommand{\emt}{\end{pmatrix}}
\newcommand{\ba}{\begin{array}{c}}
	\newcommand{\ea}{\end{array}}
\newcommand{\be}{\begin{equation}}
	\newcommand{\ee}{\end{equation}}
\newcommand{\bea}{\begin{eqnarray}}
	\newcommand{\eea}{\end{eqnarray}}

\newcommand{\bi}{\begin{itemize}}
	\newcommand{\ei}{\end{itemize}}
\newcommand{\baz}{\begin{array}{cc}}
\newcommand{\besub}{\begin{subequations}}
\newcommand{\eesub}{\end{subequations}}

\begin{document}
	\title{Cosmological Consequences of a Highly Decoupled Right-handed Neutrino}

    \author{Stephen F. King\,\orcidlink{0000-0002-4351-7507}}
	\email{S.F.king@soton.ac.uk}
	\affiliation{School of Physics and Astronomy, University of Southampton,\\ Southampton, SO17 1BJ, U.K.}

    \author{Rishav Roshan\,\orcidlink{0000-0002-8545-4188}}
	\email{r.roshan@soton.ac.uk}
	\affiliation{School of Physics and Astronomy, University of Southampton,\\ Southampton, SO17 1BJ, U.K.}
    
	\author{Sudipta Show\,\orcidlink{0000-0003-0436-6483}}
	\email{sudiptas@iitk.ac.in}
	\affiliation{Department of Physics, Indian Institute of Technology Kanpur, Kanpur 208016, India}
				
		\begin{abstract}

          In the minimal type-I seesaw, one right-handed neutrino (RHN) can decouple from neutrino
mass generation. Such a state is usually neglected, yet its feeble couplings make it
long-lived, and it can dominate the energy density of the Universe before decaying. We
study this in a minimal $U(1)_{B-L}$ model, where the RHNs are produced through $B-L$
gauge interactions, for the cases of a decoupled lightest RHN $N_1$ and a decoupled
heaviest RHN $N_3$. The resulting early matter-dominated era dilutes the baryon
asymmetry, and for decoupled $N_3$ it also delays the equilibration of the charged-lepton
Yukawa interactions, allowing a lighter $N_1$ to generate the asymmetry in the unflavoured
regime. Breaking $U(1)_{B-L}$ forms cosmic strings whose gravitational-wave background
acquires a spectral break from the early matter-dominated era, with a frequency set by the
decoupled RHN mass and the lightest neutrino mass, within reach of future detectors such as SKA, THEIA, LISA, $\mu$ARES, DECIGO and ET. 
\end{abstract}

		\maketitle

\section{Introduction}

The type-I seesaw mechanism~\cite{Minkowski:1977sc,Yanagida:1979as,Yanagida:1979gs,
GellMann:1980vs,Mohapatra:1979ia,Schechter:1980gr,Schechter:1981cv} remains the most
popular explanation for the smallness of neutrino masses. It extends the Standard Model
(SM) with right-handed neutrinos (RHNs) and, through their out-of-equilibrium decays, also
offers a route to the baryon asymmetry of the Universe via
leptogenesis~\cite{Fukugita:1986hr,Buchmuller:2005eh,Davidson:2008bu}. RHNs are
especially natural in extensions of the SM with a gauged $U(1)_{B-L}$, where they are
needed to cancel gauge anomalies and acquire their Majorana masses when the symmetry is
spontaneously broken.

Oscillation data can already be accommodated with only two RHNs taking part in the
seesaw~\cite{King:1999mb,Antusch:2011nz,Li:2017zmk,Datta:2021elq,King:2025eqv}. In this minimal
version the third state decouples from neutrino mass generation, because of a very small
Yukawa coupling, a very large mass, or both, and the lightest active neutrino is
(nearly) massless. Such a state is normally set aside as irrelevant. But the same
feeble couplings make it long-lived, and a long-lived heavy species can come to dominate
the energy density of the Universe before it decays\footnote{The cosmological role of decoupled particles is often overlooked. Recently, Ref.~\cite{Konar:2025gvh} showed that the mediator responsible for freeze-in dark matter production can dominate the energy density and induce an eMD phase, thereby affecting both dark matter production and the resulting GW signal.}.

This was demonstrated in Ref.~\cite{Roshan:2025mwl} for the heaviest RHN state $N_3$. Once
$N_3$ becomes non-relativistic it redshifts as matter and drives an early matter-dominated
(eMD) era, whose length is controlled by the same Yukawa coupling that fixes the lightest
active-neutrino mass. An important consequence is that the temperatures at which the
charged-lepton Yukawa interactions equilibrate are shifted, which can change the flavour
regime in which leptogenesis operates. However the analysis in Ref.~\cite{Roshan:2025mwl} had three limitations, as follows. It treated
only the case where $N_3$ is decoupled, although oscillation data are equally compatible
with decoupling the lightest RHN state $N_1$. It assumed that the decoupled state was produced
thermally, with a gauged $B-L$ symmetry, as mentioned only in a footnote as a way to achieve
this. And its gravitational-wave (GW) implications were discussed only qualitatively.

In this paper we address all the three points mentioned above. We work in an explicit minimal $U(1)_{B-L}$ model
with three RHNs and a $B-L$ breaking scalar. The RHNs are produced and held in
equilibrium through $B-L$ gauge interactions, and the decoupled state then falls out of
equilibrium because its Yukawa couplings are too small to keep it in contact with the SM
plasma. We study two scenarios and, in each, tie the tiny Yukawa couplings of the
decoupled state to the lightest active-neutrino mass through the Casas-Ibarra
parametrisation \cite{Casas:2001sr}.

In the case of $N_1$ being decoupled, the eMD era is driven by the lightest RHN, and leptogenesis
proceeds through the decays of $N_2$. The asymmetry is then diluted by the entropy released
when $N_1$ decays, so the observed value constrains the lifetime of $N_1$, and with it the
lightest neutrino mass. Here the shift of the charged-lepton equilibration temperatures
affects only the electron Yukawa, well below the temperature at which the asymmetry is
generated, so leptogenesis stays in the unflavoured regime.

In the case of $N_3$ being decoupled, we recover the case of Ref.~\cite{Roshan:2025mwl} within
the gauged model. The eMD era delays the equilibration of the charged-lepton Yukawa
interactions, which allows a comparatively light $N_1$ to generate the asymmetry in the
unflavoured regime, with the asymmetry again diluted by the decay of $N_3$.

The $B-L$ framework also provides a new observable. Breaking $U(1)_{B-L}$ at a high
scale forms a network of cosmic strings (CS)~\cite{Kibble:1976sj,Nielsen:1973cs}, whose GW
background would be flat in frequency if the Universe were radiation dominated throughout \cite{Auclair:2019wcv, Goncharov:2021oub,Ellis:2020ena,  Blasi:2020mfx,Samanta:2020cdk, Lazarides:2022spe, Lazarides:2022ezc, Roshan:2024qnv, Schmitz:2024gds}.
An eMD era driven by the decoupled RHN introduces a break in this plateau, after which the
spectrum falls as $f^{-1/3}$~\cite{Cui:2017ufi,Cui:2018rwi,Gouttenoire:2019kij,Blasi:2020wpy,Borah:2022byb,Borah:2022vsu,Datta:2025vyu}. The break
frequency is set by the temperature at which the eMD era ends, which scales as
$M\sqrt{m_l}$, where $M$ is the mass of the decoupled RHN and $m_l$ the lightest
active-neutrino mass. For $v_{BL}=10^{13}$~GeV ($G\mu\simeq10^{-12}$) the
resulting spectra lie within the reach of SKA, THEIA, LISA, $\mu$ARES, DECIGO and ET over a wide range
of $m_l$, down to values far below laboratory sensitivity. GW observations may therefore
probe the lightest neutrino mass and the mass of the decoupled state, linking the neutrino
sector, the thermal history and the GW signal through the same $B-L$ breaking scale. This
adds to existing proposals to probe leptogenesis and heavy-particle decays with
GWs~\cite{Dror:2019syi,Chianese:2024gee,Datta:2024tne,Murayama:2025thw}.

The paper is organised as follows. Section~\ref{secII} presents the $U(1)_{B-L}$ model.
Section~\ref{secIII} relates the Yukawa couplings of the decoupled RHN to the lightest active-neutrino
mass. Section~\ref{secIV} treats the cosmological evolution and the baryon asymmetry for the
$N_1$-decoupled and $N_3$-decoupled cases. Section~\ref{secV} computes the cosmic-string GW spectra
and their breaks. We conclude in Section~\ref{secVI}. The shift of the electron Yukawa equilibration
temperature for the $N_1$ case is given in Appendix~\ref{appendix:ET shift_N1}.

\section{Model-setup}\label{secII}

We extend the SM gauge group with an additional $U(1)_{B-L}$ gauge symmetry. In this minimal $U(1)_{B-L}$ framework, the particle content is enlarged by three RHNs, $N_1, N_2, N_3$ and a complex scalar $S$. Under $U(1)_{B-L}$, the quarks and leptons carry charges $\frac{1}{3}$ and $-1$, respectively. The quantum numbers of the additional fields under the gauge group $\mathcal{G} = SU(2)_L \times U(1)_Y \times U(1)_{B-L}$ are summarised in Table~\ref{qno}.

\begin{table}[h]
\centering
\begin{tabular}{|c c c |}
\hline
Field   & ~~~~$SU(2)_L \times U(1)_Y$ & ~~~~$Y_{BL}$ \\ \hline \hline
$N_i$ & (1,~0) & -1 \\ \hline
$S$  & (1,~0) & 2 \\ \hline
\end{tabular}
\caption{The additional fields and their quantum numbers. Here, $Y_{BL}$ denotes the $U(1)_{B-L}$ charge.}
\label{qno}
\end{table}

\noindent This choice of $B-L$ charges ensures the cancellation of the triangular $B-L$ gauge anomalies. 

Next, we write the kinetic terms for the additional fields

\begin{align}
\mathcal{L}_{KE} &=& |D_{\mu}S|^2+\sum_{i=1,2,3}\bar{N_i}i\gamma^{\mu}D_{\mu}N_i-\frac{1}{4}Z_{\mu\nu}Z^{\mu\nu},
\end{align}
with $Z^{\mu\nu}=\partial^{\mu}Z_{BL}^{\nu}-\partial^{\nu}Z_{BL}^{\mu}$, and
$D_{\mu}=\partial_{\mu}+i[Yg^{\prime}+Y_{BL}g_{BL}](Z_{BL})_{\mu}$.
\noindent Throughout, we adopt the pure $U(1)_{B-L}$ model, defined by $g^{\prime} = 0$, which forbids $Z$--$Z_{BL}$ mixing at tree level. Assuming the RHNs to be diagonal, the Yukawa Lagrangian in this setup takes the form
\begin{align}
	-\mathcal{L}= \overline{\ell_L}_{\alpha} (Y_{\nu})_{\alpha i} \tilde{H} N_{i}+ h_{ii} S  \overline{N_{i}^c} N_i+ h.c.,
\end{align}

Once the scalar field $S$ acquires a non-zero vacuum expectation value (VEV), $v_{\rm BL}$, the $B-L$ symmetry is spontaneously broken, giving rise to masses for the RHNs,
\begin{align}
M_{ii}=\sqrt{2}h_{ii}v_{\rm BL}.
\end{align}
In addition to the RHNs, the gauge boson associated with the $B-L$ symmetry also acquires a non-zero mass.

\begin{align}
  M_{Z_{\rm BL}}=2g_{\rm BL} v_{\rm BL}.
\end{align}
For the purpose of analysis, we always assume $g_{\rm BL}=1$ and the hierarchy: $M_{Z_{\rm BL}}>M_i$.

\section{Neutrino mass and Yukawa structure}\label{secIII}

After the breaking of electroweak symmetry, the light neutrino masses arise via the type-I seesaw mechanism
\begin{equation}
m_\nu\simeq m_D M_R^{-1}m_D^T,
\end{equation}

\noindent with $m_{D_{\alpha i}}=Y_{\alpha i} v/\sqrt{2}$ being the Dirac mass matrix (with $v=246$ GeV, the VEV of the SM Higgs boson and $i=1,2,3$). It is well established that the observed pattern of neutrino oscillations can be successfully explained within a framework containing only two RHNs. This scenario can be viewed as a limiting case of the conventional three-RHN seesaw model, where one RHN effectively decouples from the seesaw mechanism due to an extremely large Majorana mass, negligibly small Yukawa couplings, or a combination of both. A characteristic prediction of this minimal setup is the existence of one massless active neutrino, and it has been investigated extensively in the literature~\cite{King:1999mb,Antusch:2011nz,Li:2017zmk,King:2025eqv}. Despite this, the cosmological implications of the decoupled RHN have received little attention. In this work, we demonstrate that a decoupled RHN can induce a period of eMD, thereby significantly modifying the expansion history of the Universe.

We consider two scenarios involving a decoupled RHN: (i) the lightest RHN, $N_1$, is decoupled, and (ii) the heaviest RHN, $N_3$, is decoupled. The decoupling of these RHNs can be realized by assuming that they interact only feebly with the thermal bath after their production. Consequently, the corresponding entries of the Yukawa matrix can be parametrized in terms of the quantities $\zeta_i$ and $\epsilon_i$, respectively, with $\zeta_1,\epsilon_i\ll 1$. We then determine the allowed magnitude of these Yukawa couplings required to realize the decoupled-RHN scenarios and the resulting eMD era.

 In a minimal seesaw setup, the Yukawa interactions of the decoupled heavy RHN can be associated with the lightest active neutrino regardless of the active neutrino mass ordering, $i.e.$ normal hierarchy (NH) and inverted hierarchy (IH). The entries of $Y_{\nu}$ can be expressed using the Casas-Ibarra (CI) parametrization \cite{Casas:2001sr}:
    \begin{align}
		Y_\nu= \frac{\sqrt{2}}{v} U D_{\sqrt{m}} R^T D_{\sqrt{M}},
		\label{CI para}
	\end{align}
where $D_{m}$ and $D_{M}$ are the diagonal mass matrices for the active neutrinos and RHNs, respectively. As we demonstrate below, the Yukawa interactions of the decoupled heavy RHN can be associated with the lightest active neutrino by suitably modifying the rotation matrix $R$ of the CI parametrization. Consider the following structure of the Yukawa matrix:
\begin{align}
		Y_{\nu}^{N_1}=\left(
		\begin{array}{ccc}
		\zeta_1 & y_{e2} & y_{e3} \\
		\zeta_2 & y_{\mu 2} & y_{\mu 3} \\
		\zeta_3 & y_{\tau 2} & y_{\tau 3} \\
		\end{array}
		\right). \label{eq:yukawaN1}
	\end{align}
\begin{align}
		Y_{\nu}^{N_3}=\left(
		\begin{array}{ccc}
		y_{e1} & y_{e2} & \epsilon_1 \\
		y_{\mu 1} & y_{\mu 2} & \epsilon_2 \\
		y_{\tau 1} & y_{\tau 2} & \epsilon_3 \\
		\end{array}
		\right). \label{eq:yukawaN3}
	\end{align}
 Next, the matrices $U$ and $R$ can be written as:
\begin{align}
		U=\left(
		\begin{array}{ccc}
		U_{11} & U_{12} & U_{13} \\
		U_{21} & U_{22} & U_{23} \\
		U_{31} & U_{32} & U_{33} \\
		\end{array}
		\right), \label{UPMS}
	\end{align}
    \begin{align}
		R=\left(
		\begin{array}{ccc}
		R_{11} & R_{12} & R_{13} \\
		R_{21} & R_{22} & R_{23} \\
		R_{31} & R_{32} & R_{33} \\
		\end{array}
		\right). \label{rotation_matrix}
	\end{align}
With these matrix forms, we can derive the first (third) column of the Yukawa matrix to relate $\zeta_i$ ($\epsilon_i$) to the masses of the active neutrinos and the lightest (heaviest) RHN, $N_1$ ($N_3$). This can be achieved by comparing the Yukawa matrix in Eq. (\ref{eq:yukawaN1}) (Eq.~(\ref{eq:yukawaN3})) with Eq. (\ref{CI para}), we find that the parameters $\zeta_i$ and $\epsilon_i$ can be expressed as follows:
    \begin{align}
		\zeta_i=\frac{\sqrt{2}}{v}\sqrt{M_1}(U_{i1}R_{11}\sqrt{m_1}+U_{i2}R_{12}\sqrt{m_2}+U_{i3}R_{13}\sqrt{m_3}).	
        \label{zeta_form}
	\end{align}
    \begin{align}
		\epsilon_i=\frac{\sqrt{2}}{v}\sqrt{M_3}(U_{i1}R_{31}\sqrt{m_1}+U_{i2}R_{32}\sqrt{m_2}+U_{i3}R_{33}\sqrt{m_3}).	
        \label{epsilon_form}
	\end{align}
As a result the Yukawa matrix takes the form,
\begin{widetext}
    \begin{align}
		Y_\nu=\frac{\sqrt{2}}{v}\left(
		\begin{array}{ccc}
		\sqrt{M_1}(U_{11}R_{11}\sqrt{m_1}+U_{12}R_{12}\sqrt{m_2}+U_{13}R_{13}\sqrt{m_3}) & \# & \sqrt{M_3}(U_{11}R_{31}\sqrt{m_1}+U_{12}R_{32}\sqrt{m_2}+U_{13}R_{33}\sqrt{m_3}) \\
		\sqrt{M_1}(U_{21}R_{11}\sqrt{m_1}+U_{22}R_{12}\sqrt{m_2}+U_{23}R_{13}\sqrt{m_3}) & \# & \sqrt{M_3}(U_{21}R_{31}\sqrt{m_1}+U_{22}R_{32}\sqrt{m_2}+U_{23}R_{33}\sqrt{m_3}) \\
		\sqrt{M_1}(U_{31}R_{11}\sqrt{m_1}+U_{32}R_{12}\sqrt{m_2}+U_{33}R_{13}\sqrt{m_3}) & \# & \sqrt{M_3}(U_{31}R_{31}\sqrt{m_1}+U_{32}R_{32}\sqrt{m_2}+U_{33}R_{33}\sqrt{m_3}) \\
		\end{array}
		\right), \label{yukawa_another}
	\end{align}
  \end{widetext}  
Here, $M_1$ ($M_3$) and $m_{i(=1,2,3)}$ denote the masses of the lightest (heaviest) RHN and the active neutrinos, respectively. Finally, the entries marked by ``\#'' correspond to the second RHN, $N_2$, and can be obtained following the same procedure.

For the normal mass hierarchy, the lightest neutrino mass is identified with $m_1$. To relate the small Yukawa coupling of the $N_3$ ($N_1$), to the smallness of $m_1$, it follows from the expression for $\epsilon_i$ ($\zeta_i$) that one must require $R_{32}=R_{33}=0$ ($R_{12}=R_{13}=0$) and $R_{31}=1$ ($R_{11}=1$). Since $R$ is an orthogonal matrix, its most general form under these conditions can be written as  

\begin{align}
		R^{NH}_{N_3}=\left(
		\begin{array}{ccc}
		0 & \cos \theta_R & \sin \theta_R \\
		0 & -\sin \theta_R & \cos \theta_R \\
        1 & 0 & 0 \\
		\end{array}
		\right), \label{eq:NH}
	\end{align}

    \begin{align}
		R^{NH}_{N_1}=\left(
		\begin{array}{ccc}
		1 & 0 & 0 \\
		0 & \cos \theta_R & \sin \theta_R \\
        0 & -\sin \theta_R & \cos \theta_R \\
		\end{array}
		\right), \label{eq:NH}
	\end{align}
Here, $\theta_R=a+ib$ is a complex angle in general. In contrast, for inverted mass ordering, the lightest active neutrino mass is $m_3$. To relate the smallness of $\epsilon$ ($\zeta$) to $m_3$ for $N_3$ ($N_1$), we need $R_{31}$ ($R_{11}$) and $R_{32}$ ($R_{12}$) to be zero, and $R_{33}$ ($R_{13}$) to be 1. Given the orthogonality condition, $R_{13}$ ($R_{33}$) and $R_{23}$ will also be zero. Therefore, the structure of the rotation matrix will be:

\begin{align}
		R^{IH}_{N_3}=\left(
		\begin{array}{ccc}
		\cos \theta_R & \sin \theta_R & 0 \\
		-\sin \theta_R & \cos \theta_R & 0 \\
        0 & 0 & 1 \\
		\end{array}
		\right). \label{eq:IH}
	\end{align}

    \begin{align}
		R^{IH}_{N_1}=\left(
		\begin{array}{ccc}
		0 & 0 & 1\\
		 \cos \theta_R & \sin \theta_R & 0\\
     -\sin \theta_R & \cos \theta_R & 0 \\
		\end{array}
		\right), \label{eq:NH}
	\end{align}
The above discussion shows that the smallness of the Yukawa couplings associated with the heaviest (lightest) RHN can always be connected to the lightest active neutrino mass with an appropriate choice of $R$, regardless of the mass hierarchy (normal or inverted). Even though we do our analysis while considering normal mass ordering for active neutrinos, it is clear that our conclusions will be sufficiently general to remain valid under different mass orderings.
\section{Results}\label{secIV}
\subsection{Case I: Decoupled $N_1$}
 \paragraph{}
Once $U(1)_{\rm B-L}$ is broken, all the RHNs acquire a non-zero Majorana 
mass. Through their $B-L$ interactions, the RHNs are produced in the early 
Universe and are held in thermal equilibrium with the SM plasma. The heavier 
states $N_{2,3}$ remain in equilibrium by virtue of their sufficiently strong 
Yukawa couplings to the SM Higgs and lepton doublets. While being relativistic, the lightest state 
$N_1$, decouples from the bath once the number density of the $B-L$ 
gauge boson becomes Boltzmann-suppressed at temperatures below its mass: since 
$N_1$ couples only very feebly to the Higgs and lepton doublets, its Yukawa 
interactions are too weak to maintain equilibrium. After decoupling, $N_1$ 
remains relativistic until the temperature of the Universe drops below its 
mass, at which point it becomes non-relativistic, a scenario closely 
analogous to standard neutrino decoupling.


 The suppressed coupling simultaneously ensures that $N_1$ is sufficiently long-lived to dominate the cosmic energy density and drive an epoch of eMD, while also generating a tiny but non-zero mass for the lightest active neutrino. We denote this lightest neutrino mass by $m_l$ throughout the rest of the analysis. To achieve this, we consider the structure of the neutrino Yukawa matrix with the first column of the matrix given in Eq. (\ref{zeta_form}).

We now proceed to investigate the cosmological evolution leading to an epoch  eMD driven by the long-lived $N_1$ under the assumption that this species was in thermal equilibrium at the time of its production. The relevant dynamics are governed by the following coupled Boltzmann equations (BEs):

\begin{align}
a\frac{d\rho_{N_1}}{da}+F\rho_{N_1}=-\frac{\Gamma_{N_1}}{\mathcal{H}}\rho_{N_1} \nonumber\\
a\frac{d\rho_R}{da}+4\rho_R=+\frac{\Gamma_{N_1}}{\mathcal{H}}\rho_{N_1}
\label{BE}
\end{align}

\noindent where $\rho_R$ represents the energy density of SM radiation and $\rho_{N_1}$ and $\Gamma_{N_1}= \frac{(Y_\nu^\dagger Y_\nu)_{11}}{8\pi}M_1$ refer to the the energy density and decay width of $N_1$ respectively. The modified expansion rate can be expressed as,
\begin{equation}
\mathcal{H}^2=\frac{1}{3M_P^2}\left(\rho_R+\rho_{N_1}\right) 
\end{equation}
with $M_P=2.4\times10^{18}~\text{GeV}$ denoting the reduced Planck mass. The factor $F$ dictates the behaviour of $\rho_{N_1}$'s evolution where $F=4~(3)$ when $N_1$ is relativistic (non-relativistic). The resulting solution is then used in the relation
\begin{equation}
	T(a)=
	\left[\frac{30}{g_*(a)\pi^2}\right]^{1/4}
	\rho_R^{1/4}(a),
	\label{temp_eq}
\end{equation}
which relates the temperature of the thermal bath to the radiation energy density.

\begin{figure}[tb!]
	\centering
	\includegraphics[width=\linewidth]{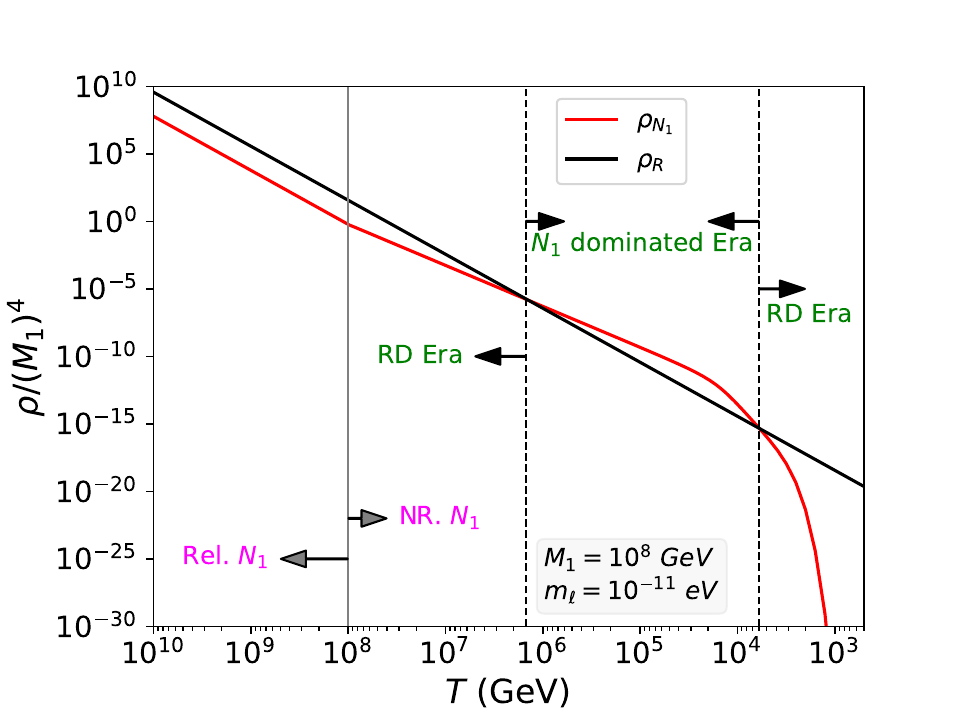}
	\caption{Evolution of the various energy densities, $\rho_R$ (black) and $\rho_{N_1}$ (red) as a function of the temperature $T$.}
	\label{fig:energy_density_N1}
\end{figure}

For illustration, in Fig.~\ref{fig:energy_density_N1}, we present a numerical solution to the above BEs showing how $N_1$ induces an era of eMD. The plot is obtained using $m_l = 10^{-11}$  eV and $M_1 = 10^{8}$ GeV. The red (black) line shows the evolution of the $N_1$ (radiation) energy density with temperature $T$. Once produced, $N_1$ behaves as radiation with energy density scaling as $\rho_{N_1} \propto a^{-4}$ and as the Universe temperature ($T$) drops to $T \sim M_1$, $N_1$ becomes non-relativistic and its energy density starts following matter-like scaling, $\rho_{N_1} \propto a^{-3}$. Consequently, the Universe enters an eMD era that persists until $N_1$ decays completely.

An immediate consequence of this $N_1$ induced eMD can be seen on the matter-antimatter asymmetry of the Universe produced from $N_2$ decay (assuming $M_1<M_2$). The asymmetry produced will then be diluted when the eMD ends and $N_1$ injects all of its energy to the thermal bath. Therefore, we proceed to evaluate the $B-L$ asymmetry using the following Boltzmann equation together with the  Boltzmann equations given in Eq.~(\ref{BE}), 
\begin{align}
a\frac{d \rho_{N_2}}{da}+F\rho_{N_2}= -\frac{\langle\Gamma_{N_2}\rangle }{\mathcal{H}}(\rho_{N_2}-\rho_{N_2}^{\text{eq}}),\nonumber
\end{align}
\begin{align}
	a\frac{d n_{\Delta}}{da}+3n_\Delta=-\frac{\langle\Gamma_{N_2}\rangle}{\mathcal{H}}\left[\frac{\varepsilon_2}{M_2}(\rho_{N_2}-\rho_{N_2}^{\rm{eq}})+\frac{n_{N_2}^{\rm{eq}}}{2 n_\ell^{\rm{eq}}}n_{\Delta}\right],
    \label{BE_asy}
\end{align}
\noindent with $\rho_{N_2}$ ( $\rho_{N_2}^{\rm eq}$) denoting the energy density (equilibrium energy density) of $N_2$, $\langle \Gamma_{N_2}\rangle$ denoting the thermally averaged decay width of $N_2$, $n_{\Delta} = n_{B-L}$ and the CP asymmetry produced from the decay of the lightest RHN $N_2$ is given by \cite{Covi:1996wh} :
\begin{equation}\label{eq:cpasym}
	\varepsilon_2=\frac{1}{8\pi} \frac{\text{Im}[({Y}_\nu^\dagger {Y}_\nu)^2_{23}]}{({Y}_\nu^\dagger {Y}_\nu)_{22}}{\cal F}\left(\frac{M_3^2}{M_2^2}\right)\,,
\end{equation}
where ${\cal F}(x)\simeq 3/2\sqrt{x}$ for hierarchical RHNs. The produced asymmetry from $N_2$ decay will be influenced by the decay of decoupled $N_1$ due to the entropy injection. The final baryon asymmetry $Y_{B}$ is then obtained by: $Y_{B}= \frac{28}{79}\sum_\alpha n_{\Delta_\alpha}/s.$

\begin{figure}[tb!]
	\centering
	\includegraphics[width=\linewidth]{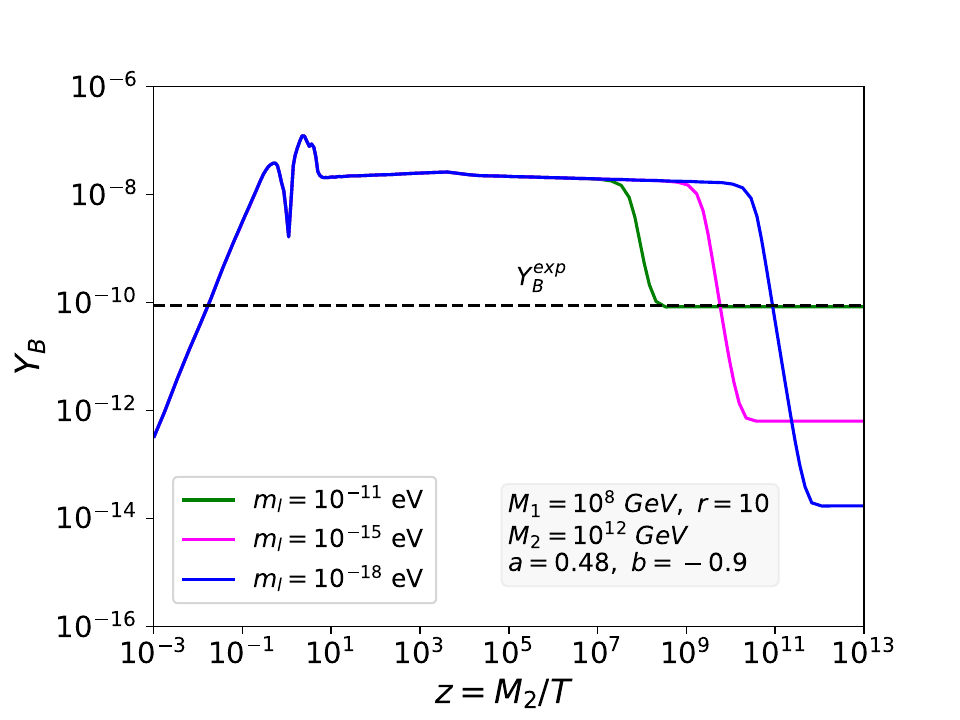}
	\caption{Evolution of the baryon asymmetry with the expansion of the Universe for a fixed value of parameters: $M_1=10^{8}~\rm{GeV},~M_2=10^{12}~\rm{GeV},~r=10,~a=0.48,~b=-0.9$ for three different values of $m_{l}$.}
	\label{YB_evo_N1}
\end{figure}

In Fig.~\ref{YB_evo_N1}, we present the evolution of the baryon asymmetry obtained by numerically solving the Boltzmann equations discussed above. As a benchmark scenario, we fix $M_2 = 10^{12}\,\text{GeV}$ and choose $r = M_3/M_2 = 10$. We further set $M_1 = 10^{8}\,\text{GeV}$ and the CI parameters to $a = 0.48$ and $a = -0.90$. We then investigate the evolution of the baryon asymmetry for three representative values of the lightest active neutrino mass, namely $m_l = 10^{-11}\,\text{eV}$, $10^{-13}\,\text{eV}$, and $10^{-15}\,\text{eV}$. The values of $a$ and $b$ are chosen such that the observed baryon asymmetry $Y_B^{\rm exp}=8.75\times 10^{-11}$ \cite{Planck:2018vyg} is satisfied for $m_l = 10^{-11}\,\text{eV}$. As seen from Fig.~\ref{YB_evo_N1}, the out-of-equilibrium decays of $N_2$ generate a lepton asymmetry, which eventually saturates once the $N_2$ number density becomes Boltzmann suppressed. The produced asymmetry remains constant and unaffected even when the Universe enters into eMD epoch induced by $N_1$. However, at the end of the eMD era, the completion of $N_1$ decays injects a significant amount of entropy into the thermal bath, thereby diluting the pre-existing asymmetry. As anticipated, smaller values of $m_l$ correspond to longer $N_1$ lifetimes, causing $N_1$ to decay later and release a larger entropy load into the plasma. As a result, the asymmetry produced from $N_2$ decays experiences a stronger dilution, leading to a substantial suppression of the final baryon asymmetry. This behaviour is clearly illustrated in Fig.~\ref{YB_evo_N1}.

\subsection{Case II: Decoupled $N_3$}
We now proceed to investigate the cosmological evolution leading to early MD driven by the long-lived heavy neutrino $N_3$ under the assumption that this species was in thermal equilibrium at the time of its production. The relevant dynamics are governed by the following coupled Boltzmann equations:

\begin{align}
a\frac{d\rho_{N_3}}{da}+F\rho_{N_3}=-\frac{\Gamma_{N_3}}{\mathcal{H}}\rho_{N_3} \nonumber\\
a\frac{d\rho_R}{da}+4\rho_R=+\frac{\Gamma_{N_3}}{\mathcal{H}}\rho_{N_3}
\label{BE_N3}
\end{align}

\noindent where $\rho_{N_3}$ and $\Gamma_{N_3}= \frac{(Y_\nu^\dagger Y_\nu)_{33}}{8\pi}M_3$ refer to the the energy density and decay width of $N_3$ respectively. In this case, the expansion rate is given by 
\begin{equation}
\mathcal{H}^2=\frac{1}{3M_P^2}\left(\rho_R+\rho_{N_3}\right) 
\end{equation}

\begin{figure}[h]
	\includegraphics[width=0.95\linewidth]{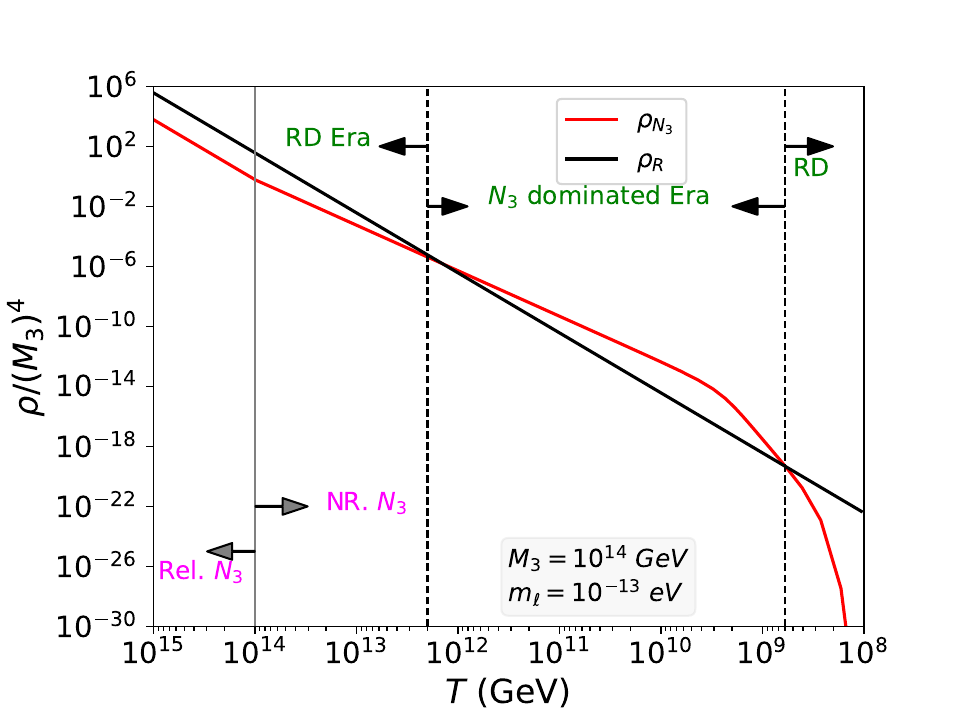}
	\caption{ Evolution of the various energy densities,$\rho_R$ (black) and $\rho_{N_3}$ (red), as a function of the temperature $T$.}
    \label{fig:energy_densityN3}
\end{figure}

Analogous to Fig.~\ref{fig:energy_density_N1}, Fig.~\ref{fig:energy_densityN3} illustrates the evolution of the energy densities of radiation and $N_3$. The results are shown for the benchmark parameters $m_l = 10^{-13}\,\text{eV}$ and $M_3 = 10^{14}\,\text{GeV}$. Similar to the behavior of $N_1$ in Case~I, the heavy neutrino $N_3$ is initially produced as a relativistic species, with its energy density scaling as $\rho_{N_3} \propto a^{-4}$. As the temperature of the Universe falls below its mass, $T \sim M_3$, $N_3$ becomes non-relativistic and its energy density subsequently evolves as $\rho_{N_3} \propto a^{-3}$. As a result, $N_3$ gradually dominates the total energy density, driving the Universe into an early matter-dominated phase. This epoch persists until the complete decay of $N_3$, after which radiation once again becomes the dominant component of the Universe.

Unlike Case~I,\footnote{While Case~I only exhibits a shift in the equilibration temperature of the SM electron Yukawa interactions, the asymmetry is generated at temperatures well above this transition. Consequently, the lepton asymmetry evolves in the unflavoured regime, rendering the final baryon asymmetry largely insensitive to flavour effects. Further details can be found in Appendix~\ref{appendix:ET shift_N1}.} the eMD phase in this scenario gives rise to an additional and phenomenologically important effect: a modification of the equilibration temperature (ET), $T^*$, of the SM charged-lepton Yukawa interactions.\footnote{A similar analysis was carried out in Refs.~\cite{Datta:2022jic, Datta:2023pav} in the context of extended reheating.} This shift can have significant implications for leptogenesis, as the equilibration of charged-lepton Yukawa interactions plays a crucial role in determining the impact of flavour effects on the generation and evolution of the lepton asymmetry \cite{Barbieri:1999ma, Nardi:2005hs, Abada:2006fw, Nardi:2006fx, Blanchet:2006be, Dev:2017trv, Datta:2021gyi}. In particular, when the Yukawa interaction of a charged-lepton flavour $\alpha$ reaches equilibrium at a temperature $T^*_{0(\alpha)}$ during a radiation-dominated (RD) era, the lepton doublets $\ell_{L\alpha}$ produced in RHN decays can efficiently interact with the corresponding right-handed charged leptons $e_{R\alpha}$ present in the thermal bath. These interactions break the quantum coherence among different lepton-doublet flavour states, thereby determining the number of independent flavour directions that must be tracked in the relevant temperature regime. 
\begin{figure}[tb!]
	\includegraphics[width=0.95\linewidth]{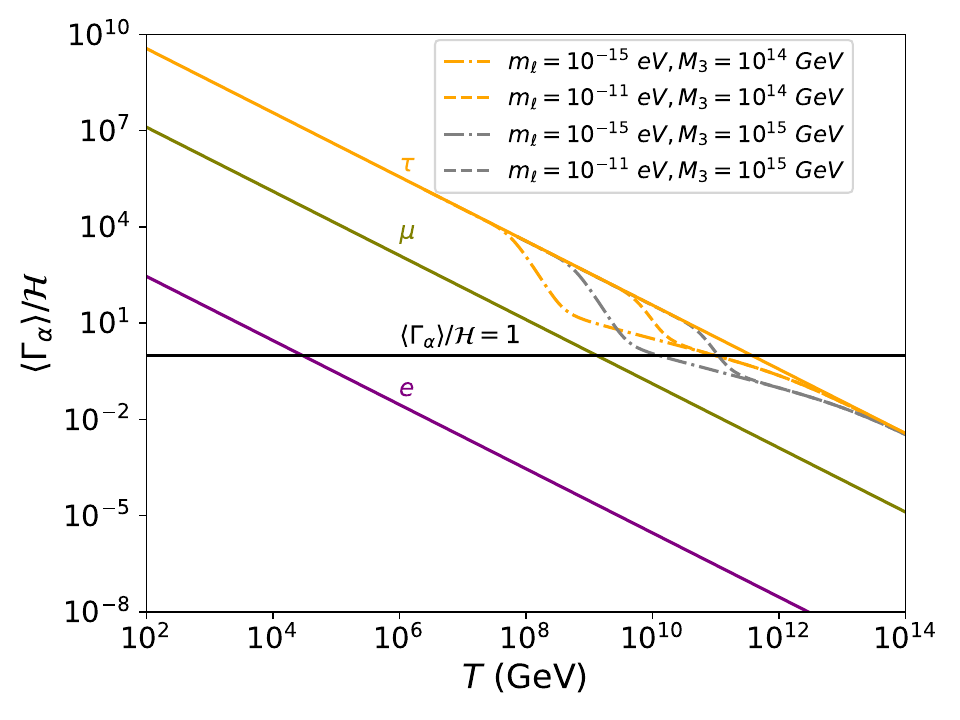}
	\caption{ Variation of $\langle \Gamma_\alpha \rangle/\mathcal{H}$ vs. $T$ for standard (solid
lines) and modified (dot-dashed and dashed) scenarios.}\label{gammabyH}
\end{figure}

\begin{figure}[tb!]
	\includegraphics[width=0.95\linewidth]{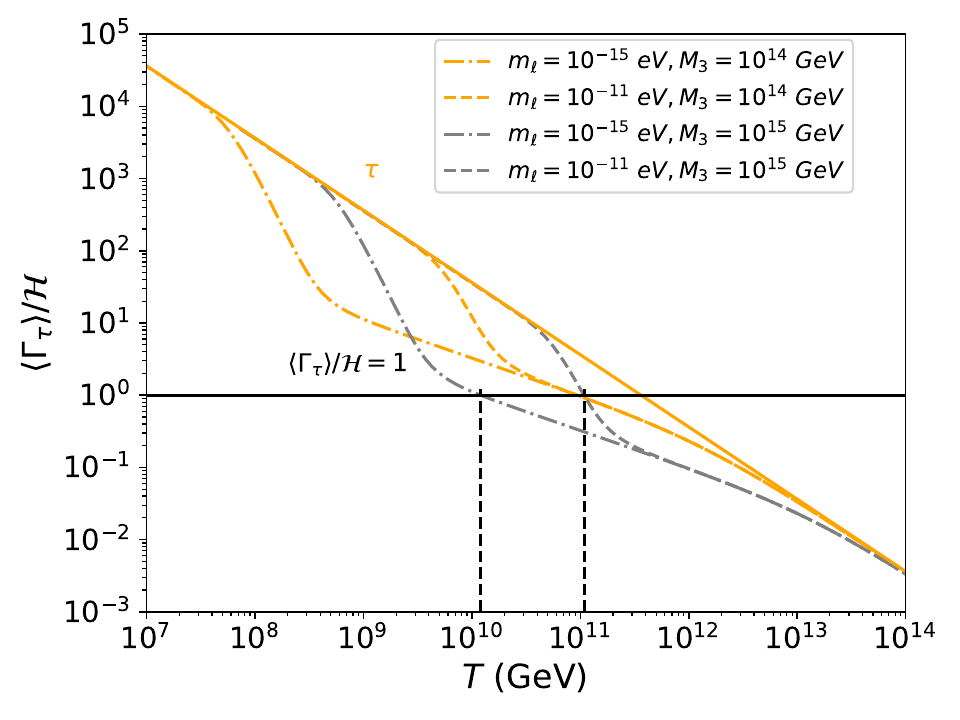}
	\caption{ Variation of $\langle \Gamma_\tau \rangle/\mathcal{H}$ vs. $T$ for standard (solid
lines) and modified (dot-dashed and dashed) scenarios.}\label{gammabyH_zoomed}
\end{figure}

In the standard RD era, the ET at which a given right-handed charged lepton, $e_{R\alpha}$, comes into equilibrium with its corresponding lepton doublet can be estimated by comparing the Higgs decay (and inverse-decay) rate, $\langle \Gamma_{\alpha} \rangle$, with the Hubble expansion rate, $\mathcal{H}$~\cite{Cline:1993vv, Cline:1993bd}. For the case of $\tau_R$, the equilibration condition in an RD Universe is given by
\begin{equation}
	\frac{\pi Y^2_{\tau}}{192 \zeta(3)} \frac{m_h^2(T)}{T}
	=
	0.066\, g_*^{1/2}\frac{T^2}{M_P},
	\label{eq-temp}
\end{equation}
where the standard RD expression
\[
\mathcal{H}(T)=0.066\, g_*^{1/2}\frac{T^2}{M_P}
\]
has been used, and $g_*$ denotes the effective number of relativistic degrees of freedom. The thermal Higgs mass is approximately given by $m_h(T)\simeq 0.6T$~\cite{Weldon:1982bn, Quiros:1999jp, Senaha:2020mop}. Solving Eq.~(\ref{eq-temp}) yields the familiar equilibration temperature
\[
T^*_{0(\tau)} = 5\times10^{11}\ {\rm GeV},
\]
which corresponds to the intersection of the solid orange curve (standard cosmology) with the horizontal black line at $\langle \Gamma_{\tau}\rangle/\mathcal{H}=1$ in Fig.~\ref{gammabyH}. Similarly, the equilibration temperatures for $\mu_R$ and $e_R$ are found to be $T^*_{0(\mu)} = 10^9\,{\rm GeV}$ and $T^*_{0(e)} = 5\times10^4\,{\rm GeV}$, respectively.

As discussed above, this usual determination of the ET implicitly assumes that the Universe remains radiation dominated throughout its evolution. In this section, we show that the presence of a eMD induced by decoupled $N_3$ can shift the ET of the charged-lepton Yukawa interactions. To distinguish these modified ETs from their standard RD counterparts, we denote them by $T^*_{\alpha}$, while retaining the notation $T^*_{0(\alpha)}$ for the values obtained in the standard cosmology. This effect is illustrated in Fig.~\ref{gammabyH} and Fig. \ref{gammabyH_zoomed}. While Fig.~\ref{gammabyH} compares the ET of different flavors in the RD era with that in the modified cosmology, Fig.~\ref{gammabyH_zoomed} focuses specifically on the $\tau$ lepton, where the shift becomes clearly visible.

The modified evolution of $\langle \Gamma_{\tau}\rangle/\mathcal{H}$, shown by the dashed and dot-dashed curves in Fig. \ref{gammabyH_zoomed}, is obtained by computing the Hubble rate from the total energy density rather than from the radiation energy density alone as in the standard scenario. The corresponding energy densities are determined by solving the Boltzmann equations in Eq.~\eqref{BE_N3}. 
The temperature of the thermal bath can then be obtained using Eq.~(\ref{temp_eq}). Owing to the contribution of $\rho_{N_3}$ to the expansion rate, the Universe evolves differently during the eMD phase, leading to a modified temperature dependence of $\mathcal{H}$. Consequently, the condition in Eq.~\eqref{eq-temp} is altered, resulting in a shift of the charged-lepton Yukawa equilibration temperatures relative to their standard RD values.

In Fig.~\ref{gammabyH_zoomed}, for $M_3=10^{14}$ GeV and $m_l=10^{-11}$ eV (dashed orange line), we observe a departure from the solid orange line. The departure begins at a high temperature when the Universe enters the $N_3$-driven matter-dominated phase and ends at a lower temperature when $N_3$ completely decays, after which the RD era resumes. Consequently, $\langle\Gamma{\tau} \rangle/{\mathcal{H}}=1$ is achieved at $T^*_{{\tau}} \simeq 10^{11}$ GeV marking a clear shift from the standard RD scenario where $T^*_{0({\tau})} = 5 \times 10^{11}$ GeV. Reducing $m_l$ to $10^{-15}$ eV (dot-dashed orange line) while keeping $M_3$ fixed does not change the initial point where $\langle\Gamma_{\tau} \rangle/{\mathcal{H}}$ departs from the standard scenario, but delays the merger with the solid orange line at lower temperature. Such a delay occurs because a smaller $m_l$ reduces the $N_3$ decay width, extending the period of the matter-dominated era. As a result, radiation domination after $N_3$ decay begins at a later epoch (lower temperature).

On the other hand, increasing the value of $M_3$ to $10^{15}$ GeV (gray dashed and dot-dashed lines) while keeping the same two values of $m_l$ increases the life-time of $N_3$, shifting the variation of $\langle\Gamma_{\tau} \rangle/{\mathcal{H}}$ to higher temperatures. In this case, the cross-over $\langle\Gamma{\tau} \rangle/{\mathcal{H}}=1$ happens at two different values of temperature: $T^*_{{\tau}} \simeq 10^{11}$ GeV for $m_l= 10^{-11}$ eV (dashed gray line) and $T^*_{{\tau}} \simeq 10^{10}$ GeV for $m_l\simeq 10^{-15}$ eV (dot-dashed gray line). The latter represents more than an order-of-magnitude shift compared to the standard scenario. Similar shifts in $T^*_{{e}}$ and $T^*_{{\mu}}$ can also be achieved by appropriately choosing $M_3$ and $m_l$. 

The modified ET has immediate implications for leptogenesis. As discussed, if the Universe is dominated by radiation as in the vanilla leptogenesis, the ET for $\tau_R,\mu_R$ and $e_R$ flavors are $T^*_{0({\tau})} = 5 \times 10^{11}$ GeV, $T^*_{0({\mu})} = 10^9$ GeV and $T^*_{0(e)} = 5 \times 10^4$ GeV, respectively. In this case, if the lightest RHN $N_1$ with mass $M_1=5\times 10^{11}$ GeV decays to produce lepton asymmetry, one needs to properly take into account the effect of the equilibration of $\tau_R$ while solving the BEs and studying the evolution of the lepton asymmetry with the expansion of the Universe. On the contrary, if the $N_1$ is heavier, $i.e~ M_1>5\times 10^{11}$ GeV, the quantum coherence between different lepton doublet states is not broken and one can stick to the unflavored leptogenesis scenario \cite{Fukugita:1986hr, Luty:1992un, Pilaftsis:1997jf, Konar:2020vuu}.

Unlike the standard leptogenesis proceeding in the RD era, an early MD driven by $N_3$ modifies the ETs as shown in Fig.~\ref{gammabyH}. With $M_1 = 5\times10^{11}$ GeV and with appropriate choices of $M_3$ and $m_l$, now the system remains in the unflavored leptogenesis regime due to the shift of $T^*_{0({\tau})}\to T^*_{{\tau}}$. Therefore, we proceed to evaluate the $B-L$ asymmetry using the following Boltzmann equation \cite{Buchmuller:2004nz}, 
\begin{align}
a\frac{d \rho_{N_1}}{da}+F\rho_{N_1}= -\frac{\langle\Gamma_{N_1}\rangle }{\mathcal{H}}(\rho_{N_1}-\rho_{N_1}^{\text{eq}}),\nonumber
\end{align}
\begin{align}
	a\frac{d n_{\Delta}}{da}+3n_\Delta=-\frac{\langle\Gamma_{N_1}\rangle}{\mathcal{H}}\left[\frac{\varepsilon_1}{M_1}(\rho_{N_1}-\rho_{N_1}^{\rm{eq}})+\frac{n_{N_1}^{\rm{eq}}}{2 n_\ell^{\rm{eq}}}n_{\Delta}\right],
    \label{BE_asy_N3}
\end{align}
\noindent with $\rho_{N_1}$ ( $\rho_{N_1}^{\rm eq}$) denoting the energy density (equilibrium energy density) of $N_1$, $\langle \Gamma_{N_1}\rangle$ denoting the thermally averaged decay width of $N_1$, and the CP asymmetry produced from the decay of the lightest RHN $N_1$ is given by:
\begin{equation}\label{eq:cpasym}
	\varepsilon_1=\frac{1}{8\pi} \frac{\text{Im}[({Y}_\nu^\dagger {Y}_\nu)^2_{12}]}{({Y}_\nu^\dagger {Y}_\nu)_{11}}{\cal F}\left(\frac{M_2^2}{M_1^2}\right)\,.
\end{equation}
Note that, in contrast to $N_3$, the RHNs $N_{1,2}$ never dominate the energy budget due to their faster decay rates and therefore do not affect ETs.

However, in a more general case where a shift of regimes of thermal leptogenesis still leads to a flavored one, the quantum transport approaches (such as the closed-time path or density-matrix formalisms) is more appropriate \cite{DeSimone:2007gkc}. In this case, the CP asymmetry $\varepsilon_{1_{\alpha}}$ (for unflavored case, $\varepsilon_{1} = \sum_{\alpha} \varepsilon_{1_{\alpha}}$) is obtained from the decay of $N_1$ to a specific flavor $\ell_{\alpha}$ and estimated using standard expression~\cite{Covi:1996wh,Nardi:2006fx,DeSimone:2007gkc,Dev:2017trv}.

\begin{figure}[h]
	\includegraphics[width=0.95\linewidth]{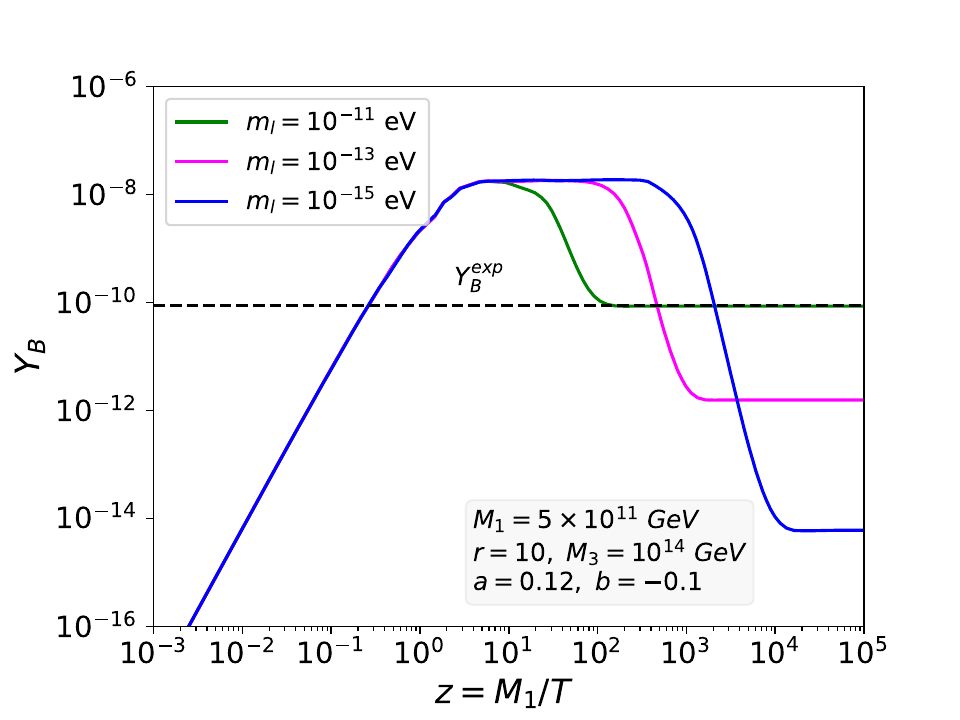}
	\caption{ Evolution of the baryon asymmetry with the expansion of the Universe for a fixed value of parameters: $M_3=10^{14}~\rm{GeV},~M_1=5\times10^{11}~\rm{GeV},~r=10,~a=0.12,~b=-0.1$ for three different values of $m_{l}$.}\label{YB_evo_N3}
\end{figure}

In Fig~\ref{YB_evo_N3}, we present the solution to the Eq. \eqref{BE_asy_N3} solved simultaneously with the BEs of Eq. \eqref{BE_N3} for a benchmark parameter set: $M_1=5\times10^{11}$ GeV, $r=10$ ($r=M_2/M_1$), $M_3=10^{14}$ GeV and $a=0.12,~ b=-0.1$. Results are shown for three values of $m_l$, with $m_l= 10^{-11}$ eV matching the observed value.  The dependence on $m_l$ is also evident: smaller values extend the $N_3$-dominated era, leading to greater dilution of the asymmetry through enhanced entropy production from $N_3$ decay. Note that, the choice presented in Fig.~\ref{YB_evo_N3} serves only as an illustrative example of how the modified expansion history delays the equilibration of the charged-lepton Yukawa interactions (shown in Fig.~\ref{gammabyH}), enabling a comparatively lighter RHN ($M_1=5\times10^{11}$ GeV) to produce the lepton asymmetry while preserving the quantum coherence among the different lepton-doublet states (unflavoured leptogenesis).

\section{Gravitational Wave signals}\label{secV}

Phase transitions involving the spontaneous breaking of a continuous symmetry, either global or gauge, can lead to the formation of topological defects known as cosmic strings \cite{Kibble:1976sj,Nielsen:1973cs,Vilenkin:1984ib,Kolb:1990vq,Preskill:1992ck,1994csot.book.....V}. Following their formation, CS evolve through intercommutation, generating closed string loops that oscillate and lose energy via radiation, predominantly in the form of GW for gauged strings and Goldstone bosons for global strings \cite{Vilenkin:1984ib,Kolb:1990vq,Preskill:1992ck,1994csot.book.....V}. In particular, gauged (local) cosmic strings constitute a well-motivated source of primordial GWs and have attracted renewed interest following the recent observation of a stochastic gravitational-wave background by the North American Nanohertz Observatory for Gravitational Waves (NANOGrav) \cite{NANOGrav:2023gor,NANOGrav:2023hvm}.

Local CS provide a unique window into high-scale physics through the stochastic GW background generated by their evolution \cite{Buchmuller:2013lra,Dror:2019syi,Buchmuller:2019gfy,King:2020hyd,Fornal:2020esl,Buchmuller:2021mbb,Masoud:2021prr,Afzal:2022vjx}. The sensitivity of GW observations to the underlying symmetry-breaking scale depends crucially on both the string dynamics and the cosmological history. In this work, we focus on local CS that are produced following the high-scale breaking of the $U(1)_{\rm B-L}$ symmetry in the early Universe. 

The evolution of a CS network is primarily characterized by the dimensionless string tension,
\begin{equation}
G\mu = \frac{v_{\rm BL}^{2}}{4M_{P}^{2}},
\label{Gmu}
\end{equation}
where $G$ is Newton's gravitational constant. Following their formation, and provided that thermal friction does not substantially damp the motion of the long strings \cite{Vilenkin:1991zk}, the network starts oscillating at a time $t_{\rm osc}$ and eventually reaches the scaling regime \cite{Blanco-Pillado:2011egf,Bennett:1987vf,Bennett:1989ak}. This regime corresponds to an attractor solution in which the growth of the long-string correlation length due to cosmic expansion is balanced by the continual production of string loops through intercommutation. The resulting loops oscillate and lose energy by emitting GW and, depending on the underlying model, particle radiation \cite{Vilenkin:1981bx,Turok:1984cn,Vachaspati:1984gt}.

The energy emitted by a loop can be decomposed into a tower of normal oscillation modes with frequencies
\begin{equation}
f_k=\frac{2k}{l},
\end{equation}
where $l$ denotes the loop length and $k=1,2,3,\ldots$ labels the harmonic mode. The total stochastic GW background is obtained by summing over the contributions from all modes,
\begin{equation}
\Omega_{\rm GW}(t_0,f)=\sum_{k}\Omega_{\rm GW}^{(k)}(t_0,f),
\end{equation}
where $t_0$ is the present time and the observed GW frequency is related to the emitted frequency through redshift,
\begin{equation}
f \equiv f(t_0)=f_k\,\frac{a(t)}{a(t_0)}.
\end{equation}
The present-day GW energy density in the $k$th harmonic mode is given by \cite{Blanco-Pillado:2013qja} \begin{align} \Omega_{\rm GW}^{(k)}(t_0,f)=\frac{2k G\mu^2\Gamma_k}{f \rho_c}\int_{t_{\rm osc}}^{t_0}dt\,\left[\frac{a(t)}{a(t_0)}\right]^5 n\!\left(t,l_k\right), \label{gwint} \end{align} where $\rho_c$ denotes the critical energy density of the Universe and $n(t,l_k)$ is the loop number density in the scaling regime. The latter can be calculated analytically within the Velocity-Dependent One-Scale (VOS) framework \cite{Martins:1996jp,Martins:2000cs,Auclair:2019wcv}\footnote{The VOS model predicts a loop number density that is approximately an order of magnitude larger than that inferred from numerical simulations. To account for this discrepancy, we introduce a normalization factor $\mathcal{F}_\alpha=0.1$, consistent with the results of Ref.~\cite{Auclair:2019wcv}.}. The quantity $\Gamma_k=\Gamma k^{-\delta}/\zeta(\delta)$ specifies the fraction of the total GW power emitted into the $k$th harmonic and depends on the small-scale structure of the loops. In particular, $\delta=4/3$ for cusp-dominated emission and $\delta=5/3$ for kink-dominated emission. Throughout this work, we assume that GW emission is dominated by cusps and therefore adopt $\delta=4/3$ in the computation of the GW spectrum.

\begin{figure}[htb!]
	\includegraphics[width=0.95\linewidth]{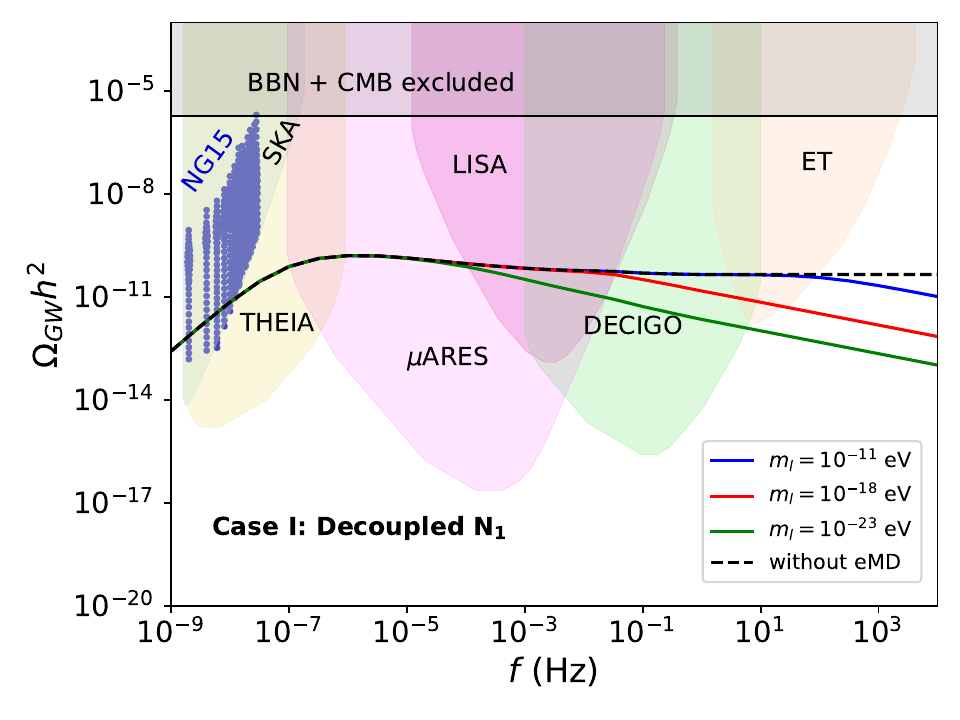}
	\caption{ The GW spectral shapes for local strings  are shown with a period of eMD for a fixed value of parameters: $M_1=10^{8}~{\rm{GeV}},~v_{\rm BL}=10^{13}{\rm{GeV}}$ for three different values of $m_{l}$.}\label{GW_caseI}
\end{figure}

A characteristic signature of the GW background sourced by cosmic strings is the presence of a nearly scale-invariant plateau, shown for example by the black dashed curve in Fig.~\ref{GW_caseI} for $f \gtrsim 10^{-5}\,\mathrm{Hz}$. This feature originates from loops that are produced and decay during the radiation-dominated era. The corresponding plateau amplitude is approximately given by 
\begin{align} \Omega_{\rm GW}^{(k=1),{\rm plateau}}(f)=\frac{128\pi G\mu}{9\zeta(\delta)} \frac{A_r}{\epsilon_r}\Omega_r \left[(1+\epsilon_r)^{3/2}-1\right], \label{flp1} \end{align}
where $\epsilon_r=\alpha/(\Gamma G\mu)$, $\alpha$ denotes the loop-size parameter at formation, $\Omega_r\simeq 9\times10^{-5}$, and $A_r=5.4$ \cite{Auclair:2019wcv}. Throughout this work, we adopt $\alpha\simeq0.1$ \cite{Blanco-Pillado:2013qja,Blanco-Pillado:2017oxo}, as suggested by numerical simulations, $\Gamma\simeq50$ \cite{Vachaspati:1984gt}, and the Cosmic Microwave Background (CMB) bound $G\mu\lesssim10^{-7}$ \cite{Charnock:2016nzm}. These values imply $\alpha \gg \Gamma G\mu$, in which case Eq.~\eqref{flp1} reduces to $\Omega_{\rm GW}^{(k=1)}\propto G\mu \propto v_{\rm BL}^2$. Consequently, models with larger symmetry-breaking scales generate stronger GW signals and are therefore more readily testable by GW experiments.

This simple plateau structure is modified if the Universe experiences an intermediate period of eMD \cite{Cui:2017ufi,Cui:2018rwi,Gouttenoire:2019kij}, which in our scenario is driven by the long-lived RH neutrino $N_1$ (case I) and $N_3$ (case II) . In that case, the spectrum develops a break at the characteristic frequency

\begin{align}
f_{\rm brk}= \sqrt{\frac{8}{\alpha\Gamma G\mu}}\, t_{\rm brk}^{-1/2} t_0^{-2/3} t_{\rm eq}^{1/6}, \end{align}

\noindent where $t_{\rm brk}$ denotes the end of the eMD era and $t_{\rm eq}$ is the time of standard matter-radiation equality. The temperature corresponding to the end of eMD, $T_{\rm brk}\equiv T_{\rm end}$, is given by 
\begin{align} 
T_{\rm end} &= \sqrt{\frac{1}{\pi}\sqrt{\frac{10}{g_*}}M_P \frac{1}{v^2} \frac{M_1^2}{\pi} \frac{m_l}{2}} ~{\rm GeV} \nonumber\\ &= 2.48\times10^{11} \left(\frac{M_1}{10^{10}\,{\rm GeV}}\right) \left(\frac{m_l}{{\rm eV}}\right)^{1/2} {\rm GeV}. 
\end{align}
The break frequency for local string networks can then be estimated from,
\begin{align}
f_{\rm brk}
&=
6.32 \times 10^{-3}\,{\rm Hz}
\left(\frac{T_{\rm end}}{\rm GeV}\right)
\left(
\frac{0.1 \times 50 \times 10^{-12}}
{\alpha \Gamma G\mu}
\right)^{1/2}\nonumber\\
&\times
\left(
\frac{g_*(T_{\rm brk})}
{g_*(T_0)}\right)^{1/4}.
\label{fbrk}
\end{align}
For frequencies above the break, $f>f_{\rm brk}$, the GW spectrum no longer remains flat and instead scales as $\Omega_{\rm GW}\propto f^{-1/3}$ when contributions from all harmonic modes are included \cite{Blasi:2020wpy,Gouttenoire:2019kij,Datta:2020bht,Gouttenoire:2019rtn}. Physically, this feature arises because loops formed around the transition from the eMD epoch to radiation domination retain information about the expansion history at the time of their formation. As a result, the change in spectral slope serves as a distinctive imprint of an eMD phase and provides a powerful probe of the long-lived RHNs population responsible for it in the present framework.

In Fig.~\ref{GW_caseI}, we display the projected sensitivity curve of future GW observatories, including ET~\cite{Punturo:2010zz}, LISA~\cite{LISA:2017pwj}, DECIGO~\cite{Kawamura:2020pcg}, $\mu$Ares~\cite{Sesana:2019vho}, SKA~\cite{Janssen:2014dka}, and THEIA~\cite{Garcia-Bellido:2021zgu}. Throughout this analysis, the sensitivity curves are derived assuming a detection threshold corresponding to ${\rm SNR}=10$ \cite{King:2023ayw}. The blue violin-shaped regions represent the NANOGrav 15-year results~\cite{NANOGrav:2023gor,NANOGrav:2023hvm}. GWs serve as an additional component of radiation and contribute to the effective number of relativistic species, denoted as \(N_{\text{eff}}\). The measurements of the CMB and Big Bang Nucleosynthesis (BBN) provide constraints on their energy density, limiting it to \(\Omega_{\rm GW}h^2 \leq 2.4 \times 10^{-6}\)~\cite{Domenech:2020ssp}, regardless of their origin, as illustrated by the gray shaded area. As a reference, we first show with a black dashed line the GW spectrum produced by oscillating cosmic-string loops in the absence of an eMD era, for a benchmark symmetry-breaking scale $v_{\rm BL}=10^{13}\,\mathrm{GeV}$ (with the corresponding value of $G\mu$ obtained from Eq.~(\ref{Gmu}). As expected for a local cosmic-string network, the spectrum rises at low frequencies and gradually approaches a nearly flat plateau at higher frequencies. The amplitude of this plateau is controlled by the symmetry-breaking scale and therefore increases with increasing $v_{\rm BL}$.

As discussed earlier, the presence of an eMD epoch significantly alters this characteristic plateau. In particular, the GW spectrum develops a break at a frequency $f_{\rm brk}$ corresponding to the end of the eMD era and hence directly related to the temperature at which the Universe returns to radiation domination. This break frequency can be estimated using Eq.~(\ref{fbrk}). The effect is clearly illustrated by the blue, red, and green curves, shown for a fixed decoupled RHN mass $M_1=10^{8}\,\mathrm{GeV}$ and different choices of the lightest neutrino mass $m_l$. As expected, smaller values of $m_l$ correspond to longer-lasting eMD periods, resulting in lower reheating temperatures and consequently a shift of the spectral break towards smaller frequencies. Conversely, larger values of $m_l$ lead to shorter eMD durations and hence larger values of $f_{\rm brk}$. For $M_1=10^{8}\,\mathrm{GeV}$, we restrict our analysis to $m_l \gtrsim 10^{-23}\,\mathrm{eV}$ in order to ensure that the end of eMD occurs before BBN, i.e., $T_{\rm end}>T_{\rm BBN}$. These benchmark serves purely as an example, qualitatively similar features arise for other choices of $v_{\rm BL}$, $M_1$, and $m_l$.

It is worth emphasizing that the shift in the GW spectrum provides not only a direct probe of an early matter-dominated epoch but also an indirect probe of leptogenesis. To illustrate this connection, we highlight a benchmark point with $m_l=10^{-11}\,\mathrm{eV}$, for which the observed order of the baryon asymmetry can be successfully reproduced (see Fig.~\ref{YB_evo_N1}). 

\begin{figure}[htb!]
	\includegraphics[width=0.95\linewidth]{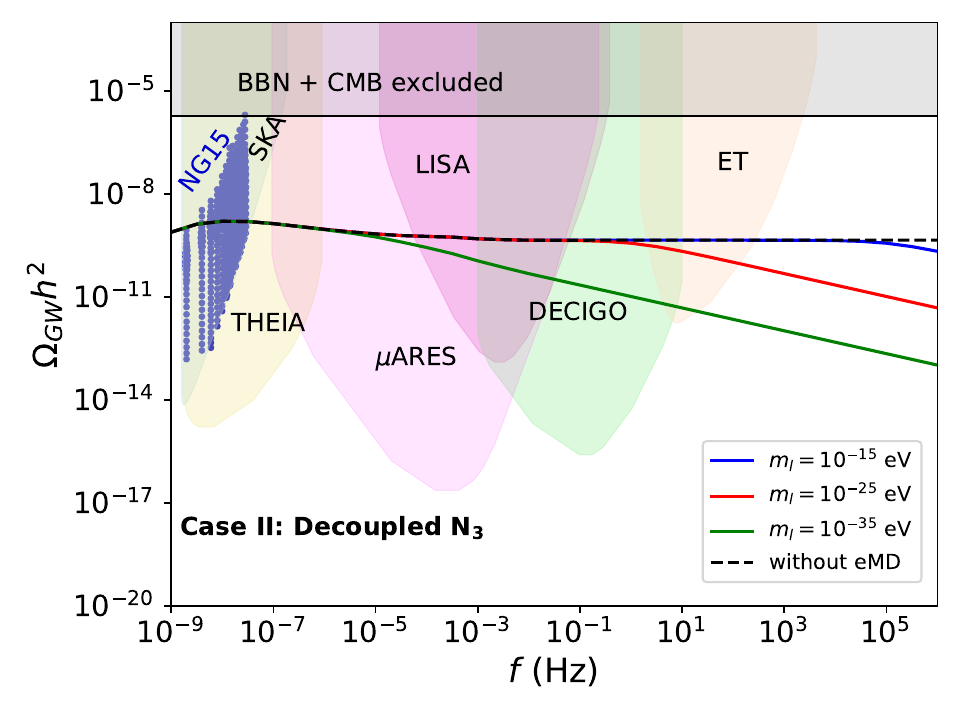}
	\caption{ The GW spectral shapes for local strings  are shown with a period of eMD for a fixed value of parameters:$M_3=10^{14}~{\rm{GeV}},~v_{\rm BL}=10^{14}{\rm{GeV}}$ for three different values of $m_{l}$.}\label{GW_caseII}
\end{figure}

Finally, in Fig.~\ref{GW_caseII}, we present the corresponding GW spectra for the scenario in which the heaviest RHN, $N_3$, decouples and drives an eMD era. We fix $M_3=10^{14}\,$GeV and accordingly set $v_{\rm BL}=10^{14}\,$GeV. For this choice of $M_3$, we restrict our analysis to $m_l \gtrsim 10^{-35}\,\mathrm{eV}$ to ensure that the eMD phase ends before the onset of BBN, i.e., $T_{\rm end}>T_{\rm BBN}$. We find that the eMD era induced by a RHN with mass $M_3=10^{14}\,$GeV can be probed by future GW experiments only if the lightest active neutrino mass satisfies $m_l \lesssim 10^{-23}\,\mathrm{eV}$. Consequently, while GW observations may serve as a probe of this eMD era, the parameter region consistent with successful baryogenesis remains inaccessible. The reason is that the associated break frequency, $f_{\rm brk}\simeq 10^{7}\,$Hz for $m_l=10^{-11}$ eV ($M_1=5\times10^{11}$ GeV), is well above the frequency reach of present and foreseeable GW experiments.

Next, we briefly comment on other potential source of GWs in the present framework. Here, gravitons can also be produced from the decay of heavy RHNs in the early Universe generating a GW signal ~\cite{Datta:2024jiv, Murayama:2025thw}. The resulting GW spectrum is typically peaked at very high frequencies. While the presence of an eMD phase would likewise modify this signal, the corresponding frequency range remains beyond the reach of current and foreseeable GW experiments. A similar high-frequency GW signature can also arise in freeze-in dark matter scenarios, where the dark matter is produced via the decay of a heavy particle~\cite{Konar:2025iuk}.

\section{Conclusion}\label{secVI}
The type-I seesaw mechanism remains one of the most compelling and extensively studied frameworks for simultaneously explaining the origin of neutrino masses and the observed baryon asymmetry of the Universe. In most studies, one of the three RHNs, typically the heaviest or the lightest, is assumed to be effectively decoupled and hence plays a negligible role in neutrino mass generation and leptogenesis. In this work, however, we demonstrate that such a seemingly inactive RHN can have a profound impact on the cosmological evolution of the early Universe. Focusing on a minimal $U(1)_{B-L}$ extension of the SM, in which RHNs are populated through $B-L$ gauge interactions, we investigate two scenarios: a decoupled lightest RHN, $N_1$, and a decoupled heaviest RHN, $N_3$. We show that the decoupled RHN can temporarily dominate the energy density of the Universe, leading to an epoch of eMD. This eMD phase has several important consequences: (i) it modifies the dynamics of lepton asymmetry generation when leptogenesis takes place during the matter-dominated era, (ii) the final baryon asymmetry is diluted through entropy injection at the end of the eMD phase, (iii) it shifts the equilibration temperatures of the charged-lepton Yukawa interactions, with the impact depending on which RHN is decoupled, and (iv) it alters the GW spectrum produced by oscillating CS loops.


Turning to the experimental prospects, we find that the observed baryon asymmetry can be successfully reproduced for light active-neutrino masses in the range $m_l \sim 10^{-11}$--$10^{-14}\,\mathrm{eV}$. This parameter space can be probed by ongoing and future direct neutrino-mass experiments, such as KATRIN~\cite{KATRIN:2021dfa} and Project 8~\cite{Project8:2017nal}. In the present set up, the breaking of the $U(1)_{B-L}$ also gives rise to CS, whose stochastic GW background exhibits a characteristic spectral break induced by the eMD phase. The resulting GW signal, including this spectral break, may be accessible to future GW observatories such as SKA, THEIA, LISA, $\mu$ARES, uDECIGO, and ET. While the spectral break provides a direct signature of the eMD epoch, it can also serve as an indirect probe of leptogenesis in the scenario with a decoupled $N_1$. In contrast, for the case of a decoupled $N_3$, the region of parameter space compatible with successful leptogenesis remains beyond the reach of the projected GW sensitivities.

\section*{Acknowledgment}			
SFK acknowledges financial support from the STFC Consolidated Grant ST/T001011/1 and Leverhulme Trust Emeritus Fellowship
Grant. SS acknowledges funding support from the Indian Space Research Organisation (ISRO) under grant STC/PHY/2024427Q.

\appendix
\section{Delayed Charged Lepton Yukawa Equilibration with decoupled $N_1$}
\label{appendix:ET shift_N1}

\begin{figure}[htb!]
	\centering
	\includegraphics[width=\linewidth]{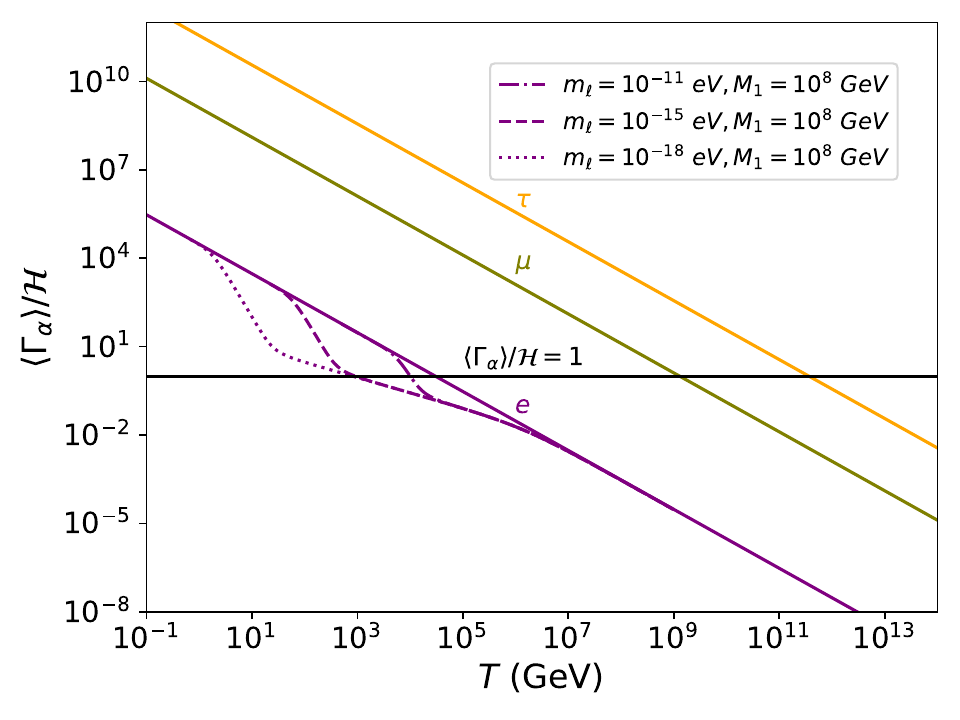}
	\caption{Variation of $\langle \Gamma_\alpha \rangle/\mathcal{H}$ vs. $T$ for standard (solid
		lines) and modified (dot-dashed and dashed) scenarios.}
	\label{delayed_ther_N1}
\end{figure}

For completeness, we examine the impact of an $N_1$-induced eMD era on the ET of charged-lepton Yukawa interactions. We consider the lightest RHN with mass $M_1=10^8\,$GeV, for which the eMD phase begins when the temperature of the Universe falls to $T\sim M_1 \simeq 10^8\,$GeV. Since the $\mu$- and $\tau$-Yukawa interactions equilibrate at higher temperatures, $T^{*}_{0(\mu)} = 10^9\,$GeV and $T^{*}_{0(\tau)} = 5\times10^{11}\,$GeV, respectively, they are already in equilibrium before the onset of the eMD era. Consequently, the eMD-induced modification of the ET can only affect the electron Yukawa interaction, as illustrated in Fig.~\ref{delayed_ther_N1}. 

For $M_1=10^8\,$GeV and $m_l=10^{-11}\,$eV, a clear deviation from the standard RD evolution (solid purple curve) is observed. This deviation starts when the Universe enters the $N_1$-dominated matter era and persists until $N_1$ completely decays, after which the standard RD evolution is restored. As a result, the condition $\langle\Gamma_e\rangle/\mathcal{H}=1$ is satisfied at $T^{*}_{e}\simeq 2\times10^{4}\,$GeV, representing a shift from the standard RD value, $T^{*}_{0(e)}=5\times10^{4}\,$GeV. 

As expected, decreasing $m_l$ while keeping $M_1$ fixed does not alter the temperature at which $\langle\Gamma_e\rangle/\mathcal{H}$ begins to depart from the standard RD prediction. However, it delays the temperature at which the ratio rejoins the standard evolution. This behavior arises because a smaller $m_l$ leads to a smaller $N_1$ decay width, thereby prolonging the duration of the matter-dominated phase. Consequently, the transition back to radiation domination occurs at a later epoch, corresponding to a lower temperature.

Although this shift in the ET of electron's Yukawa interaction has a negligible impact on the leptogenesis parameter space in scenarios with hierarchical RHNs, it can become relevant for studies of low-scale leptogenesis involving quasi-degenerate RHNs, where a proper treatment of the modified thermal history is essential.

\bibliography{ref,ref2}

\end{document}